\documentclass[letter,twocolumn]{jpsj3} 
\usepackage{ulem}
\usepackage{color}

\def\runauthor{Shin-ichi \textsc{Kimura} \textit{et al}.}
\title{
Anisotropic Electronic Structure of the Kondo Semiconductor CeFe$_2$Al$_{10}$ Studied by Optical Conductivity
}
\author{
Shin-ichi \textsc{Kimura}$^{1,2,}$\thanks{E-mail address: kimura@ims.ac.jp}, 
Yuji \textsc{Muro}$^3$,
and
Toshiro \textsc{Takabatake}$^{3,4}$
}
\inst{
$^1$UVSOR Facility, Institute for Molecular Science, Okazaki 444-8585 \\
$^2$School of Physical Sciences, The Graduate University for Advanced Studies (SOKENDAI), Okazaki 444-8585 \\
$^3$Department of Quantum Matter, ADSM, Hiroshima University, Higashi-Hiroshima, Hiroshima 739-8530 \\
$^4$Institute for Advanced Materials Research, Hiroshima University, Higashi-Hiroshima, Hiroshima 739-8530
}
\recdate{
\today
}
\abst{
We report temperature-dependent polarized optical conductivity [$\sigma(\omega)$] spectra of CeFe$_2$Al$_{10}$, which is a reference material for CeRu$_2$Al$_{10}$ and CeOs$_2$Al$_{10}$ with an anomalous magnetic transition at 28~K.
The $\sigma(\omega)$ spectrum along the $b$-axis differs greatly from that in the $ac$-plane, indicating that this material has an anisotropic electronic structure.
At low temperatures, in all axes, a shoulder structure due to the optical transition across the hybridization gap between the conduction band and the localized $4f$ states, namely $c$-$f$ hybridization, appears at 55~meV.
However, the gap opening temperature and the temperature of appearance of the quasiparticle Drude weight are strongly anisotropic indicating the anisotropic Kondo temperature.
The strong anisotropic nature in both electronic structure and Kondo temperature is considered to be relevant the anomalous magnetic phase transition in CeRu$_2$Al$_{10}$ and CeOs$_2$Al$_{10}$.
}
\kword{CeFe$_2$Al$_{10}$, Kondo semiconductor, optical conductivity, electronic structure}
\begin{document}
\maketitle
%
%
%
Recently, intricately crystallized compounds, such as those that form caged structures and triangular lattice, have attracted attention because of the novel physical properties that these structures can generate.
One of these compounds is $RM_2$Al$_{10}$ ($R$ = rare earth, $M$ = Fe, Ru, Os).
$RM_2$Al$_{10}$ has an orthorhombic YbFe$_2$Al$_{10}$-type crystal structure ($Cmcm$, $Z$~=~4), in which a triangular lattice of $R$-atoms is formed in the $bc$-plane, a square lattice in the $ac$-plane, and a hexagonal structure in the $ab$-plane.~\cite{Thiede1998}
Among these compounds, CeRu$_2$Al$_{10}$ and CeOs$_2$Al$_{10}$ have anomalous second-order antiferromagnetic phase transitions, with insulator-metal and insulator-insulator transitions, respectively, at 28~K ($T_0$).~\cite{Strydom2009,Nishioka2009,Muro2010-1,Muro2010-2,Khalyavin2010,Adroja2010,Robert2010,Kambe2010}
Because of the long distance between Ce-ions ($>$~5~\AA), the phase transition is considered to be driven by other mechanism than the Ruderman-Kittel-Kasuya-Yoshida (RKKY) interaction, which gives rise to magnetic transitions in conventional rare-earth compounds.~\cite{Matsumura2009}
Our recent study of CeOs$_2$Al$_{10}$, in which we used optical conductivity [$\sigma(\omega)$] spectra, revealed that an energy gap due to a charge instability, as well as a charge density wave (CDW), appears along the $b$-axis ($E//b$).
Along the $a$-axis ($E//a$) and $c$-axis ($E//c$), a hybridization gap, namely $c$-$f$ hybridization gap, opens between the conduction bands and nearly local $4f$ states, in the same way as in heavy fermion compounds and Kondo semiconductors.~\cite{Kimura2010}
The CDW energy gap in $E//b$ that appears at about 39~K grows with decreasing temperature, and it opens fully below $T_0$.
Observations have revealed that the change in electronic structure induces magnetic transition at $T_0$.
However, questions such as why CDW transition occurs only in $E//b$, the nature of the mechanism of the antiferromagnetic transition is related to CDW, and the nature of the interaction in CDW transition remain unanswered.
Then, we focus on a reference compound, CeFe$_2$Al$_{10}$, that has no phase transition.
It is therefore important to clarify the electronic structure of CeFe$_2$Al$_{10}$ to explore the origin of the anomalous phase transitions in CeRu$_2$Al$_{10}$ and CeOs$_2$Al$_{10}$.

CeFe$_2$Al$_{10}$ shows a Kondo effect in the electrical resistivity at high temperature and has a semiconducting behavior at low temperature.~\cite{Muro2009}
Its physical properties are similar to those of other Kondo semiconductors such as CeRhAs and CeRhSb.~\cite{Takabatake1998}
The electrical resistivity of CeFe$_2$Al$_{10}$ resembles that of CeRu$_2$Al$_{10}$ at a pressure of 4~GPa and that of CeOs$_2$Al$_{10}$ at 2~GPa.~\cite{Nishioka2009}
Therefore, CeFe$_2$Al$_{10}$ is a reference material at the higher hybridization intensities of CeRu$_2$Al$_{10}$ and CeOs$_2$Al$_{10}$.

To clarify the electronic structure of CeFe$_2$Al$_{10}$, temperature-dependent polarized $\sigma(\omega)$ spectra were measured along all principal axes.
The anisotropic electronic structure, as well as the anisotropic $c$-$f$ hybridization intensity, is discussed in this Letter.
It is found that the overall $\sigma(\omega)$ spectra of $E//a$ and $E//c$ are found to be the same as each other, but, in contrast, the spectrum in $E//b$ is very different, {\it i.e.}, the electronic structure along the $b$-axis is different from that in the $ac$-plane.
Shoulder structure due to the optical transition across a $c$-$f$ hybridization band commonly appears at $\hbar\omega$~=~55~meV in all axes, but both the gap opening temperature and the temperature of generation of the quasiparticle state at the Fermi level ($E_{\rm F}$) decrease in the order of the $b$, $a$, and $c$-axes, suggesting that the Kondo temperature ($T_{\rm K}$) is anisotropic.
These results imply that the anomalous phase transitions in CeRu$_2$Al$_{10}$ and CeOs$_2$Al$_{10}$ are related to the combination of the one-dimensional electronic structure along the $b$-axis and anisotropic $T_{\rm K}$.

%
%
Near-normal incident polarized optical reflectivity [$R(\omega)$] spectra were acquired in a very wide photon-energy region of 2~meV -- 30~eV to ensure an accurate Kramers-Kronig analysis (KKA).
Single-crystalline CeFe$_2$Al$_{10}$ was synthesized by the Al-flux method~\cite{Muro2010-2} and was well-polished using 0.3~$\mu$m grain-size Al$_{2}$O$_{3}$ lapping film sheets for the $R(\omega)$ measurements.
Martin-Puplett and Michelson type rapid-scan Fourier spectrometers 
(JASCO Co. Ltd., FARIS-1 and FTIR610) 
were used at photon energies $\hbar\omega$ of 2~--~30~meV and 5~meV~--~1.5~eV, respectively, with a specially designed feed-back positioning system to maintain the overall uncertainty level less than $\pm$0.3~\% at sample temperatures $T$ in the range of 10~--~300~K.~\cite{Kimura2008}
To obtain the absolute $R(\omega)$ values, the samples were coated {\it in-situ} with gold and then used for measuring the reference spectrum.
At $T=300$~K, $R(\omega)$ was measured for energies 1.2--30~eV by using synchrotron radiation~\cite{Fukui2001}.
In order to obtain $\sigma(\omega)$ via KKA of $R(\omega)$, the spectra were extrapolated below 2~meV with a Hagen-Rubens function,
and above 30~eV with a free-electron approximation $R(\omega) \propto \omega^{-4}$.~\cite{DG}
Electrical resistivity measurements were performed by a conventional ac four-probe method at a frequency of 19~Hz.

%
%
\begin{figure}[t]
\begin{center}
\includegraphics[width=0.37\textwidth]{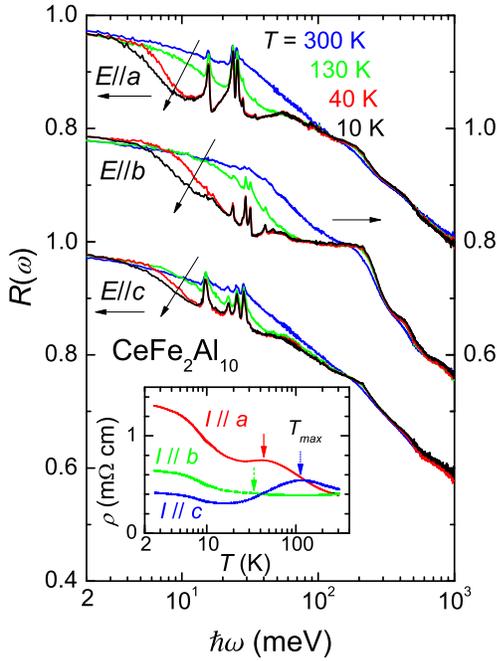}
\end{center}
\caption{
(Color online) 
The temperature-dependent polarized reflectivity [$R(\omega)$] spectra of CeFe$_2$Al$_{10}$ along the $a$-axis ($E//a$), $b$-axis ($E//b$), and $c$-axis ($E//c$).
The temperature-dependent electrical resistivity ($\rho$) along all principal axes are plotted in the inset.
The vertical arrows indicate the temperatures of the resistivity peaks in all axes.
}
\label{Reflectivity}
\end{figure}
The polarized $R(\omega)$ spectra along three principal axes are shown in Fig.~\ref{Reflectivity}.
The $R(\omega)$ at 300~K monotonically decreases with elevating photon energy up to 1~eV, which reflects the conduction band of Al expands to about 10~eV below $E_{\rm F}$ according to the band calculation.~\cite{Kimura2011}
In all polarized spectra, there are the characteristic double-peak structures in the energy range of 200--500~meV, which originates from the transition from the occupied electronic state at $E_{\rm F}$ to the unoccupied Ce~$4f$ state with spin-orbit splitting~\cite{Kimura2009-1}.
The presence of this feature indicates the stronger $c$-$f$ hybridization intensity than in CeRu$_2$Al$_{10}$ and CeOs$_2$Al$_{10}$ because the double-peak structure is not clear in these compounds.~\cite{Kimura2010,Kimura2011}

Focusing on the spectra below 100~meV reveals that there are some sharp peaks in all axes in the energy range of 10--40~meV due to optical phonons.
In comparison with the $\sigma(\omega)$ spectra of CeOs$_2$Al$_{10}$,~\cite{Kimura2010} these peaks are located at the slightly higher energy side, indicating that the phonons originate from Fe ions, and the peaks are sharper, indicating that the carrier density is smaller.
Apart from these peaks, the $R(\omega)$ spectra for all principal axes are Drude-like metallic spectra that increase to unity with decreasing photon energy.
Because the Drude-like spectral shape does not disappear even at lower temperatures, the metallic character is intrinsic.
Below 130~K, the $R(\omega)$ intensity between 5 and 100~meV rapidly decreases with decreasing temperature.
This indicates the opening of a $c$-$f$ hybridization gap.

\begin{figure}[t]
\begin{center}
\includegraphics[width=0.35\textwidth]{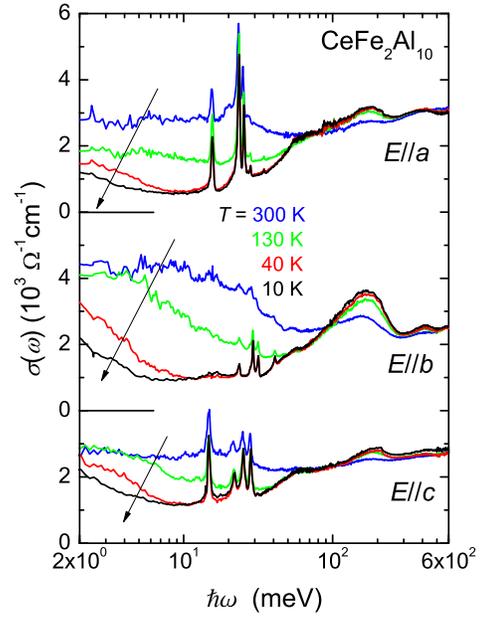}
\end{center}
\caption{
(Color online) 
Temperature-dependent polarized optical conductivity [$\sigma(\omega)$] spectra of CeFe$_2$Al$_{10}$ in the photon energy region of 2--600~meV.
The Drude weight below 50~meV at 300~K is suppressed and is shifted to the wide energy region above the $c$-$f$ hybridization energy gap at 55~meV with decreasing temperature.
}
\label{OC}
\end{figure}
The temperature-dependent $\sigma(\omega)$ spectra derived from KKA of the $R(\omega)$ spectra in Fig.~\ref{Reflectivity} are shown in Fig.~\ref{OC}.
The $\sigma(\omega)$ spectra for all principal axes at 300~K monotonically increase with decreasing photon energy and flatten below 20~meV, again indicating a metallic character.
Commonly, on all principal axes, with decreasing temperature the $\sigma(\omega)$ intensity below 50~meV decreases and a shoulder structure at 55~meV gradually evolves below 130~K.
The Drude weight at 300~K observed below 50~meV is shifted to the wide energy region above the energy gap at lower temperatures, indicating that a $c$-$f$ hybridization gap forms in a similar way as in other heavy fermion compounds and Kondo semiconductors~\cite{Kimura1994,Bucher1994,Okamura1998,Matsunami2003}.
Even at 10~K, the Drude weight remains below 10~meV, indicating the finite density of states at $E_{\rm F}$.

\begin{figure*}[t]
\begin{center}
\includegraphics[width=0.7\textwidth]{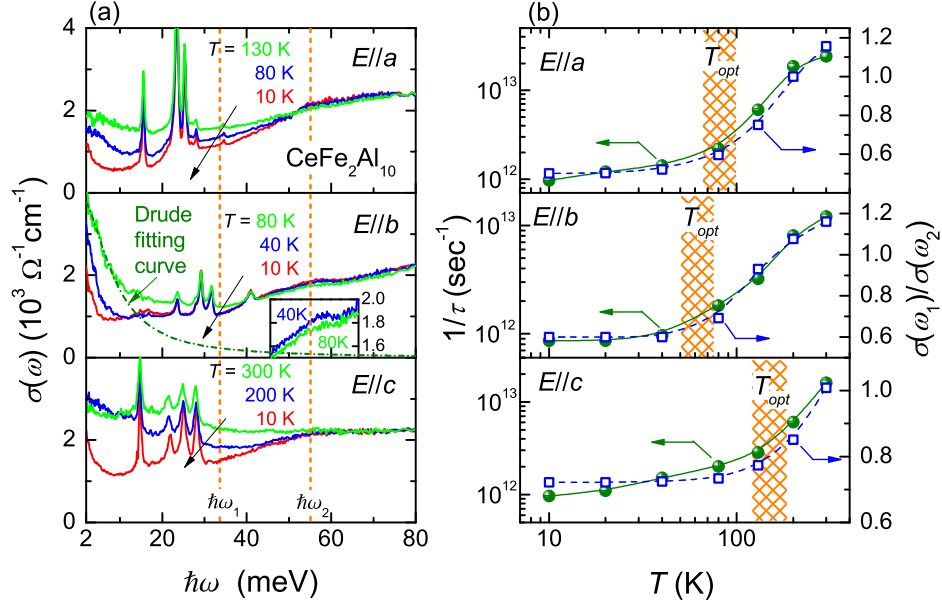}
\end{center}
\caption{
(Color online)
(a) Low-energy portion of $\sigma(\omega)$ spectra below $\hbar\omega=$80~meV in all principal axes.
One example of a Drude fitting curve for the $\sigma(\omega)$ spectrum at 80~K in $E//b$ is also plotted (dot-dashed line).
$\hbar\omega_1$ and $\hbar\omega_2$ are typical photon energies below and at the energy gap, respectively.
The inset of $E//b$ shows the enlargement near the shoulder at 55~meV.
(b) Temperature dependences of the scattering rate ($1/\tau$, solid circles) by Drude fitting and of the intensity ratio of $\sigma(\omega_1)/\sigma(\omega_2)$ (open squares).
The cross-hatchings indicate evaluated critical temperatures ($T_{opt}$) along the three principal axes.
}
\label{TdepOC}
\end{figure*}
So far, the electronic structure of CeFe$_2$Al$_{10}$ has been discussed by two groups.
One was by Chen and Lue, who used a $^{27}$Al-nuclear magnetic resonance (NMR) experiment,~\cite{Chen2010} and the other was Kawamura {\it et al.}, who used a $^{27}$Al-nuclear quadrupole resonance (NQR) experiment.~\cite{Kawamura2010}
Both groups have proposed a symmetric density of states at $E_{\rm F}$ with the band width of $140\pm40$~K (=~$11\pm3$~meV) and with the energy gap of $125\pm15$~K (=~$10\pm2$~meV).
Then a peak at $21\pm5$~meV is expected to appear in $\sigma(\omega)$ spectra.
Our $\sigma(\omega)$ spectra have a shoulder structure due to the $c$-$f$ hybridization gap at 55~meV, but no peak appears at around 21~meV.
The shoulder of 55~meV is much higher energy than $21\pm5$~meV expected by NQR and NMR.
The inconsistency between our $\sigma(\omega)$ spectra and NMR/NQR may indicate that the sizes of the indirect and direct energy gap are 10--21~meV and 55~meV, respectively.

The Drude spectral shape remains even at 10~K.
However, the $\sigma(\omega)$ intensity at the lowest accessible photon energy of 2~meV decreases to the half on cooling from 300~K to 10~K.
The temperature-dependent $\sigma(\omega)$ intensity at 2~meV is consistent with the semiconducting electrical resistivity.
The temperature dependence of the residual Drude weight is common to heavy-quasiparticles, whose scattering rate declines with decreasing temperature.~\cite{Kimura2006-1,Iizuka2010,Kimura2006-2}

Next, the anisotropic formation processes of the $c$-$f$ hybridization gap and quasiparticle state are discussed.
Figure~\ref{TdepOC}(a) indicates the temperature dependence of the low-energy portion of the $\sigma(\omega)$ spectrum along all principal axes.
The shoulder structures due to the $c$-$f$ hybridization gap commonly appear at 55~meV ($\hbar\omega_2$).
Although the gap size is similar to those of YbAl$_3$ ($T_{\rm K}\sim$670~K) and YbCu$_2$Si$_2$ ($\sim$40~K),~\cite{Kimura2009-2} the gap size is not proportional to $T_{\rm K}$.
The gap size is roughly proportional to the $c$-$f$ hybridization intensity $V$, and $T_{\rm K}\propto W_c \exp [-1/V^2D(E_{\rm F})]$, where $W_c$ is the conduction band width and $D(E_{\rm F})$ is the density of states at $E_{\rm F}$.~\cite{Hewson1993}
Therefore the different $T_{\rm K}$ implies different electronic structure.
We found a clear difference in the temperature dependence of the gap formation along the three axes.
The shoulder structure becomes visible between 130 and 80~K in $E//a$, between 80 and 40~K in $E//b$, and between 300 and 200~K in $E//c$ (Fig.~\ref{TdepOC}a).
Such anisotropic behavior has not been observed by previous optical studies of heavy fermion compounds, because few polarization measurements have been performed.

To clarify the details of the temperature dependence of the $\sigma(\omega)$ spectra, the value [$\sigma(\omega_1)/\sigma(\omega_2)$] of the $\sigma(\omega)$ intensity at 33~meV ($\hbar\omega_1$) below the shoulder divided by that on the shoulder at $\hbar\omega_2$, and the scattering rate ($1/\tau$) derived from a fitting of the Drude formula [$\sigma(\omega)=\sigma_{DC}/(1+\omega^2\tau^2)$, where $\tau$ is the relaxation time] are plotted as a function of temperature (Fig.~\ref{TdepOC}b).
The curve of $\sigma(\omega_1)/\sigma(\omega_2)$ is very similar to that of the logarithmic $1/\tau$ on each axis.
This means that the temperature dependence of $1/\tau$ is governed by $c$-$f$ hybridization gap formation, {\it i.e.}, quasiparticles due to the hybridization grow up on cooling.
These values commonly flatten below 80~K in $E//a$, below 50~K in $E//b$, and below 150~K in $E//c$, as indicated by $T_{opt}$ in the figure.
$T_{opt}$ is similar to the temperature of the electric resistivity peak in the inset of Fig.~\ref{Reflectivity}, {\it i.e.}, 45, 35, and 110~K in $I//a$, $I//b$, and $I//c$, respectively.
Because of the Kondo-like behaviors in the resistivity above the temperatures, the peaks originate from the Kondo effect.
Therefore $T_{opt}$ is also related to $T_{\rm K}$.
This result is consistent with our previous optical studies of heavy fermion compounds.~\cite{Kimura2006-1, Iizuka2010, Kimura2006-2}
Therefore, the anisotropic $T_{opt}$ implies that $T_{\rm K}$ is also anisotropic.

The similar anisotropic $T_{\rm K}$ has been discussed in orthorhombic CeNiSn and CeRhAs, in which no magnetic phase transition takes place.~\cite{Takabatake1998}
In CeOs$_2$Al$_{10}$ and CeRu$_2$Al$_{10}$, however, an anomalous antiferromagnetic transition appears due to a charge instability as well as CDW along the $b$-axis.
Along the $b$-axis, the electronic structure is one-dimensional and the $T_{\rm K}$ is lower than that in the $ac$-plane.
This is expected from the crystal structure because the Ce and Fe ions are connected in the $ac$-plane.~\cite{Muro2010-2}
The CDW transition is considered to emerge due to the combination of the one-dimensional electronic structure as well as the anisotropic $T_{\rm K}$.

\begin{figure}[t]
\begin{center}
\includegraphics[width=0.35\textwidth]{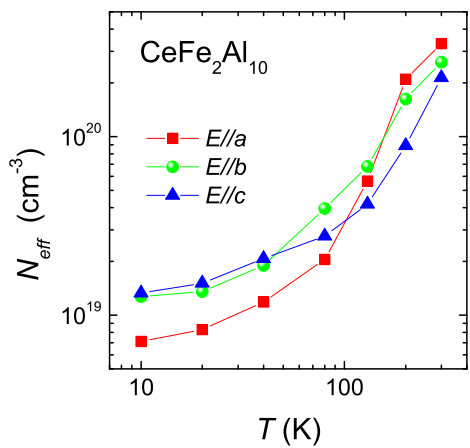}
\end{center}
\caption{
(Color online)
Temperature-dependent effective carrier density ($N_{eff}$) evaluated from the Drude fitting of the optical conductivity spectra in all principal axes.

}
\label{Neff}
\end{figure}
Finally, the origin of the semiconducting electrical resistivity in CeFe$_2$Al$_{10}$ is discussed.
The $R(\omega)$ as well as the $\sigma(\omega)$ spectra shown in Figs.~\ref{Reflectivity} and \ref{OC} do not commonly indicate semiconducting, but metallic ground state.
At low temperature, a small Drude weight from the quasiparticle state is observed, but the intensity is not large.
The Drude fitting of the $\sigma(\omega)$ leads to the temperature-dependent effective carrier density [$N_{eff}=(\sigma_{DC}\cdot m_0)/(\tau\cdot e^2)$, where $m_0$ is the free electron mass and $e$ is the elementary charge].
As shown in Fig.~\ref{Neff}, $N_{eff}$ decreases by more than one order of magnitude from 300 to 10~K in all axes, although the electrical resistivity increases only a few times.~\cite{Muro2009}
According to $N_{eff}=N_0\cdot m_0/m^*$, where $N_0$ is an actual carrier density and $m^*$ is the effective mass of carriers, the semiconducting electrical resistivity is due to either the decrease of $N_0$ or the increase of $m^*$.
Since the electronic specific heat coefficient $\gamma$ is 14~mJ/(mol$\cdot$K$^2$), which is smaller than that of the non-magnetic reference LaFe$_2$Al$_{10}$, $m^*$ should not be largely enhanced at low temperature.~\cite{Muro2009}
Therefore, the semiconducting character is attributed to the low carrier density at low temperature.
This implies that a degenerate semiconducting or semi-metallic electronic structure with a quasiparticle state with low scattering rate is realized in CeFe$_2$Al$_{10}$.
This result is consistent with the Korringa relation in the NQR data below 20~K.~\cite{Kawamura2010}

%
%
In conclusion, temperature-dependent polarized optical conductivity [$\sigma(\omega)$] spectra of CeFe$_2$Al$_{10}$ along all principal axes were measured to resolve the anisotropic electronic structure.
The shoulder of a $c$-$f$ hybridization gap of 55~meV is commonly observed along the three axes, but the electronic structure along the $b$-axis is very different from the electronic structure in the $ac$-plane.
The observed Kondo temperature is strongly anisotropic, that along the $b$-axis is the lowest.
The density of states at the Fermi level decreases on cooling, but a quasiparticle state with low scattering rate remains even at 10~K.
The observed anisotropies in the electronic structure and in the Kondo temperature are considered to be related to the anomalous antiferromagnetic phase transition of CeOs$_2$Al$_{10}$ and CeRu$_2$Al$_{10}$. 

We would like thank Professors Sera and Nishioka for their useful comments and discussions.
Part of this work was performed by the Use-of-UVSOR Facility Program (BL7B, 2010) of the Institute for Molecular Science.
The work performed in Okazaki was partly supported by a Grant-in-Aid for Scientific Research (B) (Grant No.~22340107) and for Priority Areas (Grant No.~20102004) from MEXT of Japan.
%

%
\end{document}